\documentclass[12pt]{iopart}
\usepackage{iopams}  

\usepackage{graphicx}
\usepackage{adjustbox}
\usepackage{epsfig}
\usepackage{subfigure}
\usepackage[table,xcdraw]{xcolor}
\usepackage{multirow}
\usepackage[normalem]{ulem}
\useunder{\uline}{\ul}{}
\usepackage[
singlelinecheck=false % <-- important
]{caption}
\usepackage{tikz,xcolor,hyperref}

\definecolor{lime}{HTML}{A6CE39}
\DeclareRobustCommand{\orcidicon}{%
	\begin{tikzpicture}
	\draw[lime, fill=lime] (0,0) 
	circle [radius=0.16] 
	node[white] {{\fontfamily{qag}\selectfont \tiny ID}};
	\draw[white, fill=white] (-0.0625,0.095) 
	circle [radius=0.007];
	\end{tikzpicture}
	\hspace{-2mm}
}

\foreach \x in {A, ..., Z}{%
	\expandafter\xdef\csname orcid\x\endcsname{\noexpand\href{https://orcid.org/\csname orcidauthor\x\endcsname}{\noexpand\orcidicon}}
}
\begin{document}
	
	\title[Bone Segmentation in Contrast Enhanced WB-CT]{Bone Segmentation in Contrast Enhanced Whole-Body Computed Tomography}
	
	\author{Patrick Leydon\textsuperscript{1,4}\orcidA{},Martin O'Connell\textsuperscript{1,3}, Derek Greene\textsuperscript{2}\orcidC{} and Kathleen M Curran\textsuperscript{1}\orcidB{}}

	\vspace{5pt}
	\address{\textsuperscript{1} School of Medicine, UCD, Belfield, Dublin, Ireland}
	\address{\textsuperscript{2} School of Computer Science, UCD, Belfield, Dublin, Ireland}
	\address{\textsuperscript{3} Mater Misericordiae University Hospital, Dublin, Ireland}
	\address{\textsuperscript{4} Department of Applied Science, LIT, Limerick, Ireland}
	%\ead{patrick.leydon@lit.ie}
	\vspace{5pt}
		%Required for preprint
%	\begin{indented}
%		\item \textbf{\textit{This work has been submitted to The Journal of Physics in Medicine \& Biology for possible publication. Copyright may be transferred without notice, after which this version may no longer be accessible.}}
%	\end{indented}
	\begin{indented}
		\item[]August 2020
	\end{indented}
	
	\begin{abstract}
	Segmentation of bone regions allows for enhanced diagnostics, disease characterisation and treatment monitoring in CT imaging.
	In contrast enhanced whole-body scans accurate automatic segmentation is particularly difficult as low dose whole body protocols reduce image quality and make contrast enhanced regions more difficult to separate when relying on differences in pixel intensities.
	
	This paper outlines a U-net architecture with novel preprocessing techniques, based on the windowing of training data and the modification of sigmoid activation threshold selection to successfully segment bone-bone marrow regions from low dose contrast enhanced whole-body CT scans. The proposed method achieved mean Dice coefficients of 0.979 $\pm$0.02, 0.965 $\pm$0.03, and 0.934 $\pm$0.06 on two internal datasets and one external test dataset respectively. We have demonstrated that appropriate preprocessing is important for differentiating between bone and contrast dye, and that excellent results can be achieved with limited data.		
\end{abstract}

%
% Uncomment for keywords
\vspace{2pc}
\noindent{\it Keywords}: segmentation, bone, computed tomography, contrast enhanced, low dose.
%
% Uncomment for Submitted to journal title message
%\submitto{\JPA}
%
% Uncomment if a separate title page is required
%\maketitle
% 
% For two-column output uncomment the next line and choose [10pt] rather than [12pt] in the \documentclass declaration
%\ioptwocol
%

\section{Introduction}

\subsection{Contrast Enhancement and WB-CT Bone Segmentation}\label{subsec:bckgrnd}

In oncology, once metastasis from the primary tumour site has occurred, or in bone specific cancers such as Multiple Myeloma, the cancer may manifest anywhere in the skeletal system meaning only a scan along the entire patient volume will ensure all potential sites are captured \cite{hillengass2019international,hoilund2018cancer,macedo2017bone}.

The ability to automatically isolate the bone-bone marrow (BBM) from the original scan allows for quicker, more reliable diagnosis, enhanced therapies, interventions and monitoring, as well as progressing the overall clinical understanding of a condition through advanced analytics \cite{kim2019visual} and insight into disease pathology \cite{gordon2008automated}.

Manual segmentation of a Whole Body Computed Tomography (WB-CT) would be too time consuming and a fast, accurate method of segmenting the BBM has long been a focus of research in CT imaging. Initial research focused on traditional image processing methods such as watershed \cite{boehm1999three}, level sets / graph / cuts \cite{pinheiro2015new,krvcah2011fully}, deformable models  \cite{burdin1994surface,burnett2004deformable,franzle2014fully}, self organizing maps \cite{natsheh2010segmentation,guo20183d} and others \cite{sharma2010automated,puri2012semiautomatic}. Human bones can take on a wide variety of shapes, sizes, and compositions ranging from long bones, such as the femur to irregular bones found in the vertebral column or the skull. This makes the task of WB segmentation particularly difficult when relying on any single traditional image processing method.

More recently, Deep Learning (DL) techniques, specifically Convolutional Neural Networks (CNNs), have offered a solution to this complex problem, and represent the method of choice for those with the access to large image datasets, and the resources needed to label, train and test a CNN model \cite{belal2019deep,klein2019automatic,sanchez2020segmentation,noguchi2020bone}.

Contrast Enhanced (CE) CT is routinely employed in clinical practice as a means of improving the soft tissue visibility in a scan through the introduction of a contrast agent into the body, such as iodine or barium \cite{boehm2016physics}. 
The relatively high density of contrast agents has the effect of increasing the Hounsfield Unit (HU) of the region in which they localize \cite{lusic2013x}. 
From a segmentation perspective, the presence of contrast in a scan makes it more difficult to isolate BBM from a region where contrast has localized as the HU will fall in the range for BBM \cite{fiebich1999automatic}.
An inverse relationship between CT tube voltage and change in HU has been identified \cite{kalra2019contrast} meaning the effects of contrast are greater for low voltage scan protocols, such as WB-CT. As such, segmentation of BBM in CE WB-CT is a challenging task, where CNN-based techniques are particularly useful.

\subsection{CNNs for Segmentation}\label{subsec:cnns_seg}

The main mechanism on which these networks operate relies on producing a \textit{feature map} from the convolution operation that is carried out, along with activation and pooling. This is applied at several stages in a network, with a feature map generated at each stage. Higher levels of abstraction are achieved with each successive layer, that is from lines and edges to texture to specific objects \cite{goodfellow2016deep2}. 
In the past decade, the improved computational efficiency of GPU processing, in combination with the  availability of data, and the development of high level software for development, has seen widespread application of CNNs to solving medical imaging tasks such as disease classification, object localisation, image registration, and semantic segmentation \cite{litjens2017survey}. 

The fully convolutional design of Ronneberger et al's U-net \cite{ronneberger2015u} has been widely adopted by the research community as the architecture of choice for medical image segmentation tasks. This design introduced the \textit{skip-connection} as part of an encoder-decoder network to enable the original spatial resolution of the input image to be retained, without the computational burden of previous patch-based approaches \cite{ciresan2012deep,drozdzal2016importance}.

This design has demonstrated strong performance with limited training data \cite{iglovikov2017satellite} and has been used for organ segmentation \cite{kazemifar2018segmentation}, bone segmentation \cite{iglovikov2018paediatric,zeng20173d,klein2019automatic,sanchez2020segmentation,noguchi2020bone}, lesion detection \cite{christ2017automatic}, and has also been modified to incorporate the 3-D nature of many medical imaging modalities \cite{milletari2016v,cciccek20163d}.

\section{Materials and Methods}
The following section will provide details relating to compilation of the training/testing dataset including additional HU ranging preprocessing steps, as well as a description of the U-net architecture used with associated hyperparameters. The technique for selection of the threshold for the Sigmoid Activation output is also described as well as details of the metrics for assessing the U-net models.
\subsection{Training Data}\label{subsec:TrainingData}
Once ethical approval from the institution was granted a database of 11 Positron Emission Tomography-CT (PET-CT) scans (male $= 6$ and female $= 5$) was collated and anonymised. Study inclusion criteria required patients to be over the age of eighteen and to have no active underlying conditions that would effect the BBM appearance on the scan. This was verified through review of referrals, clinical notes and the radiology reports. 

Patients had a mean age of $55 \pm16$ years and a mean weight of $77 \pm12$ kgs.  PET-CT data was collated as part of a larger study but only the CT component was used for BBM segmentation.

All scans were performed on the same Siemens Biograph 16 PET-CT scanner between May 2014 and March 2017 using the same low dose WB helical scan protocol with a 0.98 mm\textsuperscript{-1} pixel spacing, at 100 kVp, 512 x 512 size, and a  1.5 mm slice thickness.  Seven of the scans were from the skull to the proximal femur, and three were from the skull to the feet. The mean number of slices per scan was $718 \pm187$ slices. All scans were CE through the use of oral and/or intravenous (IV) dye.

It has been observed that labelling every slice is an inefficient approach for volumetric medical data, since neighbouring images contain practically identical information \cite{cciccek20163d}. In order to account for this, each patient dataset was subsampled to select every 5\textsuperscript{th} slice. 
This reduced the amount of manual review/correction, while still retaining a large enough selection of images to sufficiently capture the anatomical variance of a WB scan.  
\begin{figure}[htbp]
	\centering
	\includegraphics[width=1\textwidth]{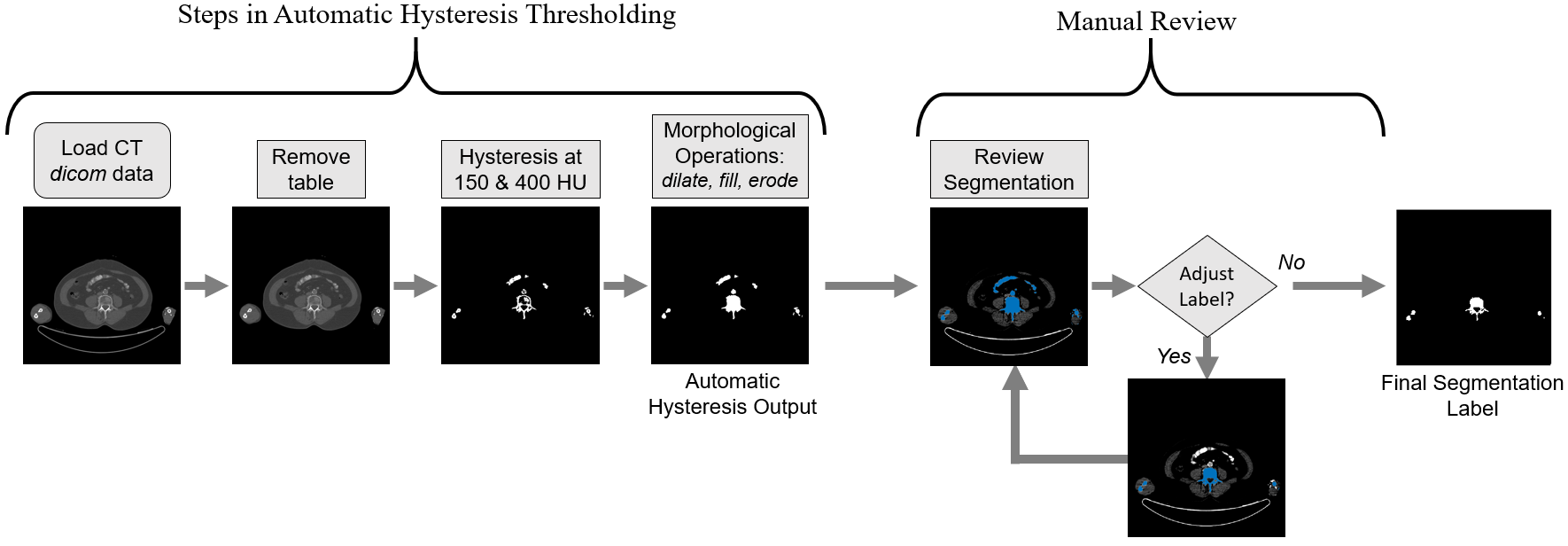}
	\caption{Workflow for training data labelling. The steps resulting in the hysteresis output are fully automated. The segmentation labels, shown in blue, are reviewed and manually corrected if needed.}.
	\label{fig:workflow_label}
\end{figure}
\subsubsection{Training Data Labelling:}\label{subsubsec:TrainingDataLabelling}

A typical single WB-CT scan may have over 900 individual axial slices to capture the entire patient volume, each of which will require a corresponding label for CNN training. The gold standard is manual delineation by a clinical expert \cite{klein2019automatic,suzuki2010computer,tappeiner2019multi,alirr2018automated,noguchi2020bone}, however given the substantial quantity of data required for deep learning, it is unfeasible to apply entirely manual methods in labelling of BBM segmentation images.

A solution to labelling vast numbers of images is to first apply a standard automated approach such as hysteresis thresholding \cite{canny1983finding} with HU thresholds set at $T_{upper} = 400$ and $T_{lower} = 150$ \cite{sogo2012assessment}. This  is followed by a series of morphological dilation, erosion, and fill operations \cite{eddins2004digital} to capture the low density marrow regions. Finally, if any corrections are required this is done as part of a manual review by an expert, see Figure \ref{fig:workflow_label}. Matlab \cite{MATLAB:2018b} was used for all data labelling steps, as well as subsequent steps in training/testing dataset creation, as described in Section \ref{subsec:windowing_images}.  %The initial steps are the automated methods of BBM segmentation which will yield an output in which may, or may not, need correction.
\begin{figure}[htbp]
	\centering
	\subfigure[The head support and the region filling within the skull needs to be manually removed from BBM label.]{\includegraphics[width=0.23\textwidth]{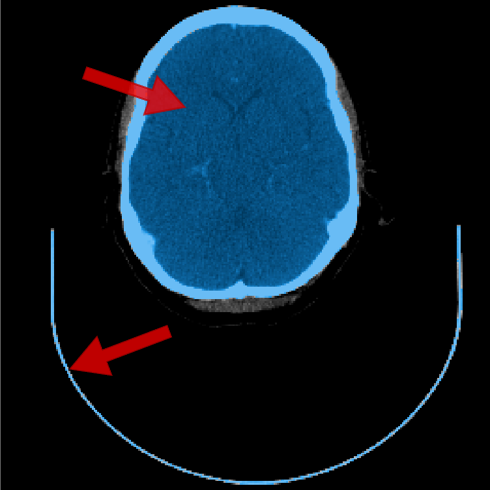} \label{subfig:hys_seg_error1}}
	\hspace{0.1cm}
	\subfigure[Iodine contrast is within the HU range of hysteresis thresholding and requires correction in the final BBM label.]{\includegraphics[width=0.23\textwidth]{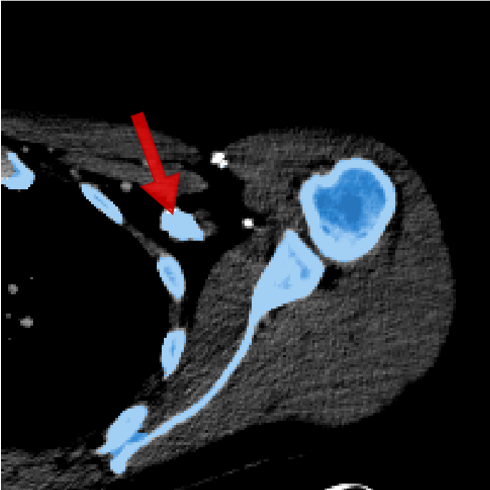} \label{subfig:hys_seg_error2}}
	\hspace{0.1cm}
	\subfigure[Metal streak artifacts are included in hysteresis segmentation as they are in the upper HU range.]{\includegraphics[width=0.23\textwidth]{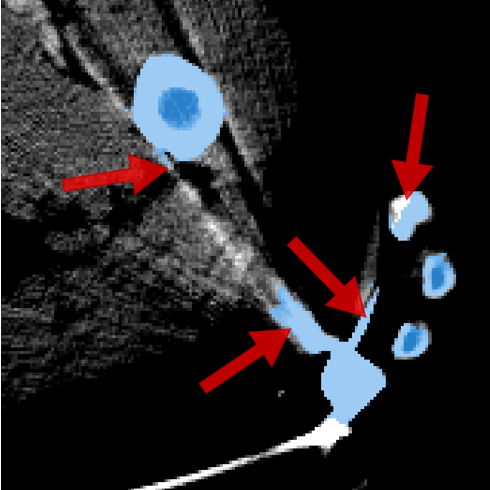} \label{subfig:hys_seg_error3}}
	\caption{The hysteresis output segmentation is overlayed in light blue on the input CT image. Example of errors are indicated by the red arrows.} 
	\label{fig:Errors_hysteresis}
\end{figure}

There were typical shortcomings of hysteresis thresholding for BBM segmentation which were evident from review and required correction. Firstly, the choice of $T_{upper}$ and $T_{lower}$ results in retention of the patient table and head support which were removed via methods similar to those outlined in \cite{zhu2012automatic,bandi2011automated}, see Figure \ref{subfig:hys_seg_error1}.
Secondly, a morphological fill operation was necessary in order to capture lower HU marrow regions in the segmentation. However, this results in the skull vault and the spinal cord being filled and included in the segmentation, see Figure \ref{subfig:hys_seg_error1}.
Thirdly, the use of contrast dye \cite{baron1994understanding} means that soft tissues that would normally be outside the BBM hysteresis threshold range are included in the segmentation, see Figure \ref{subfig:hys_seg_error2}.
Finally, as shown in Figure \ref{subfig:hys_seg_error3}, prominent high density streak artifacts due to the presence of metallic objects \cite{barrett2004artifacts} are retained in hysteresis thresholding.

Despite the need for manual correction in the previously mentioned circumstances, this approach provides labelled data in sufficient numbers for DL applications, while requiring substantially less user input in comparison to entirely manual delineation of images.
The final training dataset consisted of 1574 Axial CT images of size 512 x 512 as well as the corresponding BBM binary labels.
\subsubsection{Internal Testing Data:}
In addition to the training data, a further two patient datasets were randomly selected for testing, corresponding to an approximate 15:85 split for testing and training by image count. The test scans took place between October 2013 and November 2014, and were carried out on the same PET-CT scanner using the same low dose WB protocol as the training data. 

It is essential that the separation of data for training and testing is based on patient scans due to the large degree of correlation that exists between image slices in close proximity \cite{cciccek20163d}. The standard deep learning approach of a random dataset shuffle followed by a training, validation, and testing split \cite{willemink2020preparing} is unsuitable for medical volumetric data, and may produce misleading test results.

The data was prepared in the same fashion as previously described in Section \ref{subsubsec:TrainingDataLabelling} yielding a final internal test dataset consisting of 228 Axial CT images of size 512 x 512 and the corresponding BBM binary labels.
Both of the internal test datasets were from skull to proximal femur and both were CE scans, see Figure \ref{fig:dsc_images}.
\subsubsection{External Testing Data:}
In order to further validate performance and assess generalisability the U-net BBM segmentation models were also tested on an external dataset. The dataset has been made publicly available by Peréz-Carrasco \textit{et al} \cite{perez2018joint}. In their research they have analysed bone segmentation using energy minimisation techniques, and the dataset consists 27 Axial small CT volumes in 20 patients, with a total of 270 images and corresponding BBM labels available.

An additional benefit to using this external test dataset is that it has also been used by other researchers \cite{klein2019automatic,noguchi2020bone} in their own versions of the U-net for bone segmentation. This allows for a direct comparison to these studies to be included as part of this research. It is worth noting that, unlike other studies, we have not used any of the external dataset for training. Instead we have retained it for testing purposes only.
\subsubsection{Preprocessing of Training and Testing Images:}\label{subsec:windowing_images}
\begin{figure}[htbp]
	\centering
	\subfigure[Original DICOM across full HU range with contrast dye visible in the stomach.]{\includegraphics[width=0.30\textwidth]{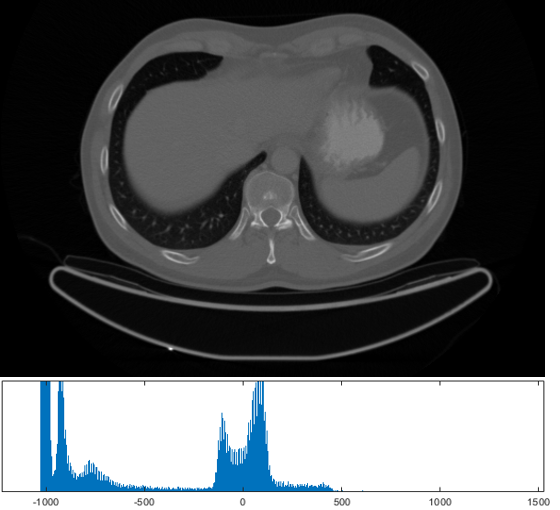} \label{subfig:convertHU1}}
	\hspace{0.2cm}
	%\subfigure[-100 to 1500 HU.]{\includegraphics[width=0.25\textwidth]{images/convert_1500}} \label{subfig:convertHU2}
	%\hspace{0.2cm}
	\subfigure[Dense cortical bone is clearly distinguishable from contrast at -100 to 1000 HU range.]{\includegraphics[width=0.30\textwidth]{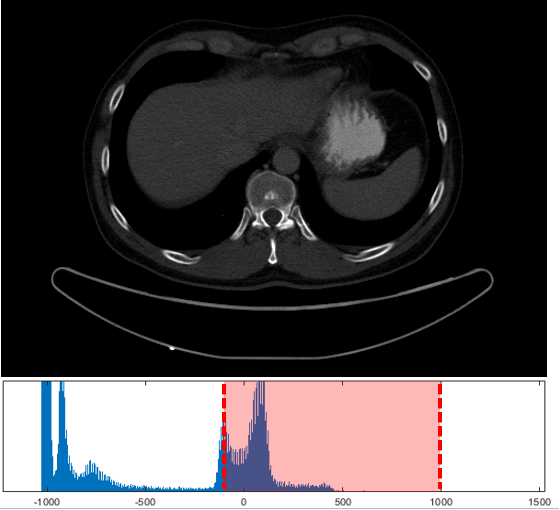} \label{subfig:convertHU3}}
	\hspace{0.2cm}
	%\subfigure[-100 to 750 HU.]{\includegraphics[width=0.25\textwidth]{images/convert_750}} \label{subfig:convertHU4}
	%\hspace{0.2cm}
	%\subfigure[-100 to 500 HU.]{\includegraphics[width=0.25\textwidth]{images/convert_500}} \label{subfig:convertHU5}
	%\hspace{0.2cm}
	\subfigure[At \textit{default} range of 0 to 255 HU cortical bone and contrast dye are compressed to same intensity value.]{\includegraphics[width=0.30\textwidth]{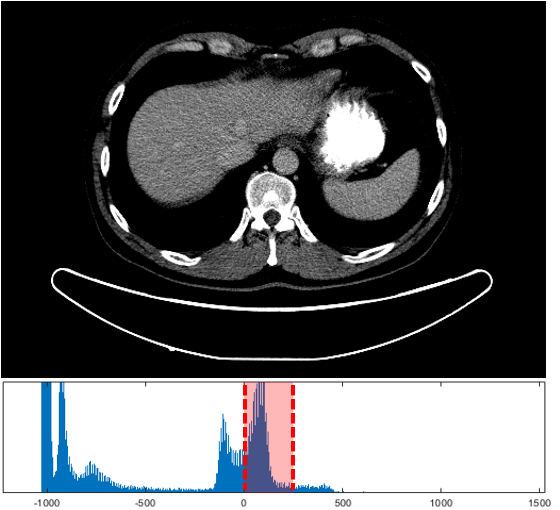} \label{subfig:convertHU6}}
	%\caption[to be completed.]
	\caption{Example slice, and associated histograms, from the Internal Test Dataset 1 demonstrating the impact of HU range on visibility of bone and contrast dye.}
	\label{fig:convertHU} 	
\end{figure}
When converting from the 12-bit grayscale DICOM to an 8-bit PNG file, Matlab will compress the image automatically. Once the CT calibration factor for conversion to HU has been applied, this produces an image between 0 and 255 HU. This means that any tissues below 0 or above 255 HU are assigned those values and much of the contrast detail relating to bone, bone marrow as well as contrast dye is lost, see Figure \ref{subfig:convertHU6}. Whereas when a larger range is used, such as -100 to 1000 HU as shown in Figure \ref{subfig:convertHU3}, the dye is more identifiable by pixel intensity alone.

For training and testing purposes, five separate HU ranged datasets were created from the DICOM files: -100 to 1500, -100 to 1000, -100 to 750, -100 to 500, and the default of 0 to 255. These ranges were selected to capture different amounts of BBM, and contrast dye detail \cite{sogo2012assessment}.
\subsection{Training of U-net Model}\label{subsec:Ch5_Unet_training}
\subsubsection{Augmentation: }\label{subsec:preprocessing}
The network architecture used here was based on the original U-net design \cite{ronneberger2015u}, and is a \textit{fully} convolutional design with skip connections. The network was implemented in Python using Keras with Tensorflow v2.0.
\begin{figure}[htbp]
	\centering
	\includegraphics[width=0.9\textwidth]{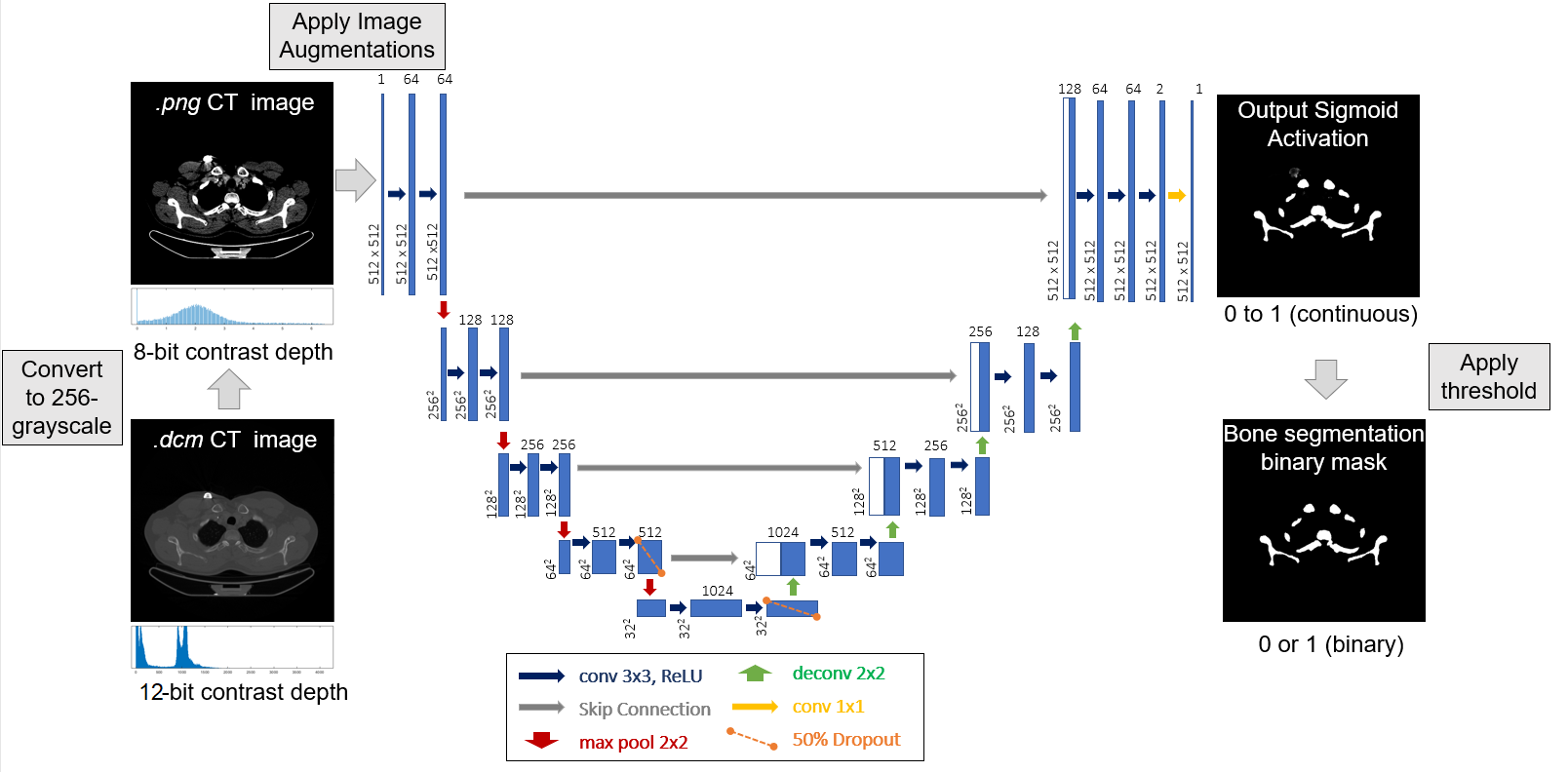}
	\caption{The U-net architecture \cite{ronneberger2015u} modified for CT bone segmentation.}
	\label{fig:UNet_flow2}
\end{figure}
Additional preprocessing augmentations were applied to improve generalisability as has been done in \cite{ronneberger2015u,klein2019automatic}. This consisted of rotation $\pm 2^{\circ}$, height/width shifts $\pm 5\%$, shear $\pm 5\%$, zoom $\pm 5\%$, and horizontal flipping. The degree of augmentation was randomly allocated within the ranges indicated.

The network consisted of a contracting path of 14 layers of downsampling convolutions and an expansive path of 14 layers of upsampling deconvolutions with pooling with ReLU activations \cite{nair2010rectified} using a stride of two at each layer and max pooling, as shown in Figure \ref{fig:UNet_flow2}.

The Adam optimizer \cite{kingma2014adam,reddi2019convergence,goodfellow2016deep2} with a learning rate of $\eta = 1\times10^{-5}$ \cite{feng2017discriminative} and a drop-out of 50\% was applied. Binary Cross Entropy (BCE) was used as the loss function for training \cite{drozdzal2016importance}.

Training was performed for 100 epochs, each having a batch size of two, with 300 steps per epoch. The choice of batch size was limited by GPU memory constraints. The literature suggests that a larger batch size, as in \cite{klein2019automatic}, is preferable. 
A 80:20 split was applied to the training data to retain a validation dataset to fine tune the hyper-parameters and verify that the model was not overfitting.

\subsection{Sigmoid Activation Threshold} \label{subsec:Ch5_unet_thres_sel}
The final layer of the U-net (Figure \ref{fig:UNet_flow2}) is the output from a sigmoid activation function that produces a map of continuous values between zero and one. In order to convert this to a functional binary mask, a threshold must first be applied. In previous work this has been set to a fixed value of 0.5 \cite{ibtehaz2020multiresunet,leger2019deep,ding2019votenet}.

The approach used in this paper to determine the threshold for the sigmoid activation output was based on finding an optimum balance between true positive (TP) and false positive (FP) rates based on the precision-recall curve for a range of threshold choices between zero and one.
The optimum threshold will be that which corresponds to best balance of precision and recall, determined by $\min\nolimits{\mid 1-(\frac{recall}{precision})\mid}$. This process was applied for each of the U-net models using the training and validation data.
\subsection{Segmentation Analysis} \label{subsec:Ch5_Analysis}
The Dice Similarity Coefficient ($DSC = \frac{2 TP}{2TP + FP + FN}$)  \cite{dice1945measures} is an overlap-based metric \cite{taha2015metrics} that we apply to assess the segmentation outputs from the models. This was performed separately for each image to assess performance across the various anatomical regions within the testing datasets. %Means and standard deviation (sd) are presented for all models on each test dataset in the results section.

To determine the significance of differences in performance between the models, analysis of variance (ANOVA) \cite{mathworks2019statistics} was applied to model DSC scores for Internal Dataset 1, 2, and the External Dataset. DSC scores tended towards a value of one, indicating that the results are left skewed. ANOVA requires that the data follow a normal distribution, and so a logit transform was applied to the DSC results prior to ANOVA testing, where $logit(DSC) = ln(\frac{DSC}{1 - DSC})$ \cite{zou2004statistical}. This produces an approximately normal distribution suitable for ANOVA. The null hypothesis  here is that there is no difference between the models, and is rejected only if a \textit{p}-value of $<0.05$ is returned.

\section{Results}
%
%	There is a brief initial description of model behaviour during the training process followed by details of the sigmoid activation thresholding based on the P-R curve method. 
In this section we present results for the HU ranged model segmentations on test datasets.  To demonstrate the performance variation across patient anatomy, the DSC scores on individual axial slices of Internal Datasets 1 and 2 are also presented. Example images have been included to illustrate differences between models as well as other challenges specific to the data. We have also compared our results with other similar bone segmentation studies.

%\subsection{Indications from Training}\label{subsec:Ch5_Unet_training_findings}

%There were no signs of overfitting as the validation losses followed similar patterns and values as the training losses for all models. The largest reduction in losses occurs within the first 25 epochs for all models with subsequent incremental reductions observed beyond this point.

%All of the BCE HU ranged models achieved similar performance in the final Mean Square Error, Loss and Accuracy reached during training with \textbf{-100 to 1500}, \textbf{-100 to  500}, and \textbf{0 to 255} being marginally superior to \textbf{-100 to 1000}, and \textbf{-100 to 750}.
%
%
%
\subsection{Sigmoid Activation Threshold Results}\label{subsec:Threshold_sel_res}
\begin{table}[htbp]
	\caption{Optimal threshold values for U-net sigmoid activation layer of HU range models from precision-recall curves of Training Data.}
	\begin{center}
		\begin{adjustbox}{max width=\textwidth}
			\begin{tabular}{lccccc}
				\hline
				\textbf{HU Range}  & \textbf{-100 to 1500} & \textbf{-100 to 1000} & \textbf{-100 to 750} & \textbf{-100 to 500} & \textbf{0 to 255} \\ \hline
				\textbf{Threshold} & 0.42                  & 0.57                  & 0.59                 & 0.41                 & 0.45              \\ \hline
			\end{tabular}
		\end{adjustbox}
	\end{center}
	\label{table:sig_thres_window}
\end{table}
The sigmoid activation threshold selections, based on the precision-recall curve of the training and validation datasets, are presented in Table \ref{table:sig_thres_window}. 
All of the HU range models demonstrated different optimum threshold levels in relation to the standard value of 0.5, with the closest being 0.45 for the 0 to 255 HU model. 

For the other models, larger differences between our approach and the standard threshold were observed. Our results indicate that the application of a standard threshold of 0.5 would negatively impact final binary masks, which highlights the importance of this factor, particularly for -100 to 1500, -100 to 1000, -100 to 750, and -100 to 500 HU ranged models.
\subsection{Dice Similarity Coefficient Results}
DSC scores for models ranged between of $0.979\pm0.021$ and $0.921\pm0.069$ for -100 to 1000 on Internal Dataset 1 and 0 to 255 on the External Dataset, see Table \ref{table:DSCresWindow}. The DSC results for Internal Dataset 1 were higher than Internal Dataset 2 and the -100 to 500 a model demonstrated the lowest overall performance on internal data. On the External Dataset the -100 to 500 and 0 to 255 models were the best and worst performing achieving DSC scores of $0.934\pm0.059$ and $0.921\pm0.069$ respectively.

\begin{figure}[htbp]
	\centering
	\includegraphics[width=.9\textwidth]{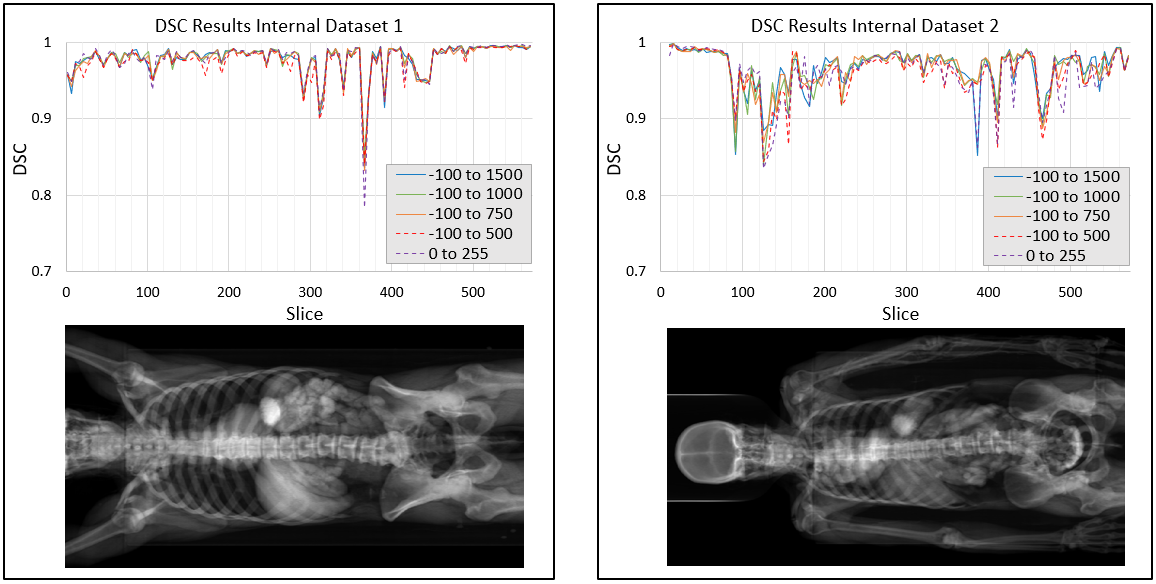}
	\caption{DSCs for individual axial slices on Internal Datasets 1 and 2 demonstrating performance of different HU ranged models across patient volumes. Grayscale coronal projections of respective CT volumes have been included as position references.}
	\label{fig:dsc_images}
\end{figure}

The ANOVA results demonstrated that there were no significant differences between models in the Internal Datasets. For the External Dataset a \textit{p}-value of 0.04 was observed between the -100 to 500 and 0 to 255 models indicating a significant difference between these models.
\begin{table}[htbp]
	\caption{DSCs ($\pm$ sd) for models trained on images over various HU ranges with BCE loss.}
	\begin{center}
		\begin{adjustbox}{max width=\textwidth}
			\begin{tabular}{rccccc}
				\hline
				\multicolumn{1}{l}{}  & \textbf{Internal Data 1}  &  & \textbf{Internal Data 2}  &  & \textbf{External Data}    \\ \hline
				\textbf{H.U. Range}   & \textbf{DSC ($\pm$ sd)} &  & \textbf{DSC ($\pm$ sd)} &  & \textbf{DSC ($\pm$ sd)} \\
				\textbf{-100 to 1500} & 0.978 $\pm$ 0.022         &  & 0.965 $\pm$ 0.028         &  & 0.923 $\pm$ 0.067         \\
				\textbf{-100 to 1000} & 0.979 $\pm$0.021          &  & 0.965 $\pm$ 0.028         &  & 0.922 $\pm$ 0.070         \\
				\textbf{-100 to 750}  & 0.978 $\pm$ 0.022         &  & 0.965 $\pm$ 0.026         &  & 0.933 $\pm$ 0.052         \\
				\textbf{-100 to 500}  & 0.975 $\pm$ 0.022         &  & 0.959 $\pm$ 0.031         &  & 0.934 $\pm$ 0.059         \\
				\textbf{0 to 255}     & 0.979 $\pm$ 0.025         &  & 0.959 $\pm$ 0.033         &  & 0.921 $\pm$ 0.069         \\ \hline
			\end{tabular}
		\end{adjustbox}
	\end{center}
	\label{table:DSCresWindow}
\end{table}
As our internal test data are whole-body (WB) scans, it is useful to plot the DSC scores for individual axial slices across the entire patient volume. This is presented in Figure \ref{fig:dsc_images} with projection images included as positional references. Both of the graphs reveal similar patterns for all HU ranged models, with corresponding high and low scores occurring in approximately the same anatomical locations.

From this we can see that Internal Dataset 1 is generally consistent across the volume with the main reduction in DSC scores localized within vertebrae of the lower spine where models have partially included intervertebral discs in their segmentations. It is evident that models appear most stable in the thorax and proximal femurs. All of the models either removed all or the majority of the contrast dye for this test dataset. Only the -100 to 1500  and 0 to 255 HU models retained small regions of contrast in the stomach in a single slice however a DSC score of 0.989 for both was recorded for this image indicating excellent overlap with ground truth.

The main source of segmentation error not associated with contrast dye was due to an implanted cardiac device with no model successfully removing the device in its entirety in the final segmentation. It is worth noting that no patients in the training data had such a device, see examples in bottom row of Figure \ref{fig:results_examples1}.

The plot of DSC scores across volume of Internal Dataset 2, shown in Figure \ref{fig:dsc_images}, demonstrates a higher degree of variability both between models and across the patient volume. The main factors impacting performance in all models was the presence of metal streak artifacts due to high levels of dental filling, as well as jewellery, and unique to this patient was the presence of a ring pessary device in the pelvis. There was a large amount of contrast dye present in this scan which proved more challenging from a segmentation perspective, most notably IV contrast in the arm, subclavian arteries, and descending aorta that was partially included in several segmentations, see Figure \ref{fig:results_examples2}.

\subsection{Segmentation Examples}
Examples of the segmentations for a selection of the HU ranged models on each of the test datasets is presented in the following section, as part of a visual assessment of segmentations to demonstrate where models were found deviate, or where stand-out features were observed. For comparison, WB projections of BBM segmentations have also been included alongside the initial hysteresis technique used as part of the data labelling workflow. Due to the small volumes in the external dataset such projections were performed for the internal datasets only.

\subsubsection{Internal Test Dataset 1:}

\begin{figure}[htbp]
	\centering
	\includegraphics[width=0.85\textwidth]{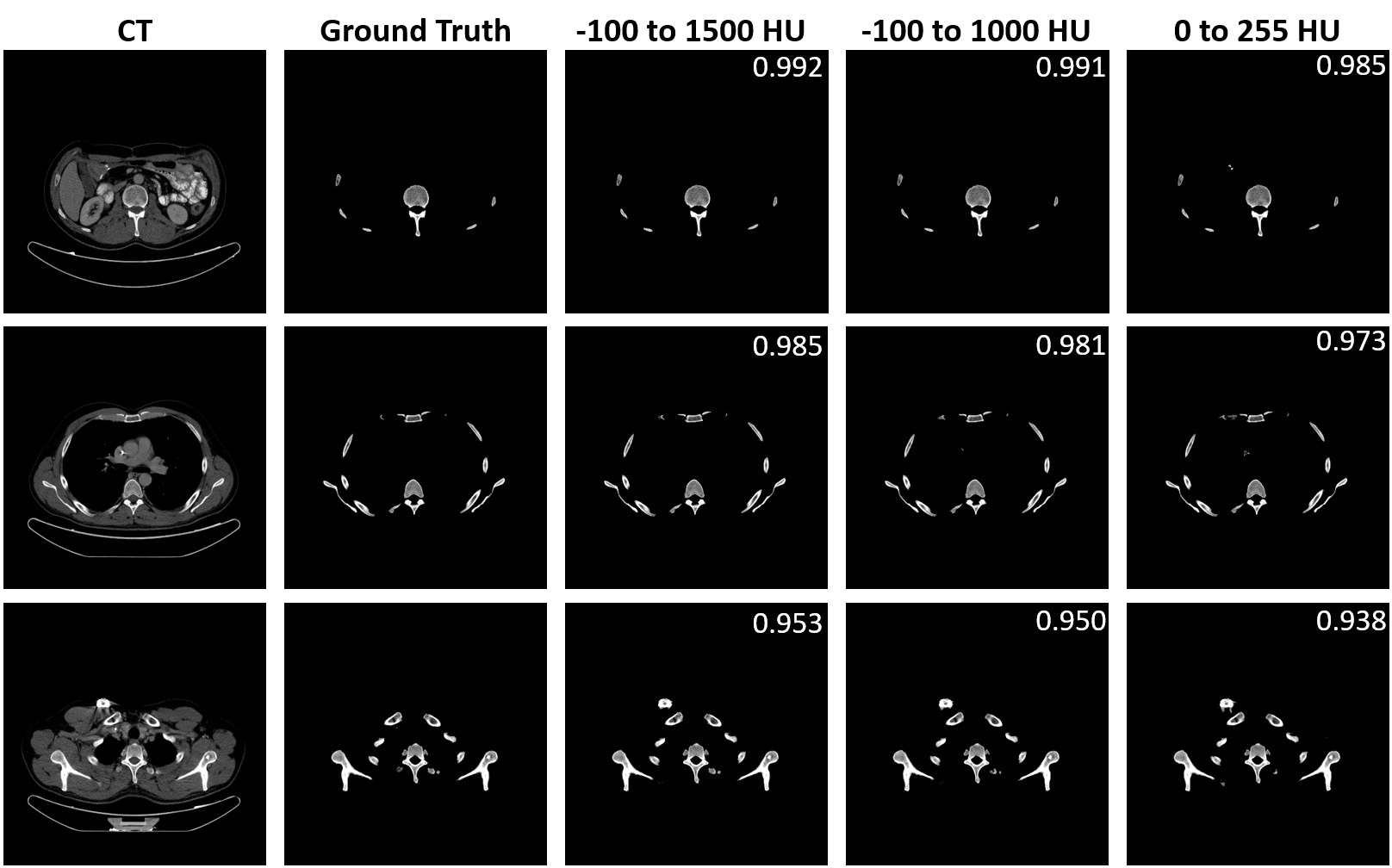}
	\caption{A selection of CT inputs, ground truth, and U-net segmentations demonstrating a selection of DSC results, provided in top right corners, on Internal Test Dataset 1.}
	\label{fig:results_examples1}
\end{figure}

From Figure \ref{fig:results_examples1} we can see that the models trained on -100 to 1500 and -100 to 100 HU successfully removed all of the contrast in the abdomen, and the 0 to 255 model retained a small portion, with respective DSC scores of 0.992, 0.991, and 0.985 recorded. The metal cable of the pacemaker device, present in a total of 23 images of Internal Dataset 1, was completely removed by the -100 to 1500 HU model, whereas intermittent removal was noted in other models. The -100 to 500 HU model demonstrated the worst performance in terms of removal of the metal cable, with partial inclusion on the final segmentation noted on 17 images. The housing of the cardiac device, located on the anterior chest wall was the main source of error for this dataset with progressively more streak artifacts included in final segmentations as the HU range was reduced, see bottom row Figure \ref{fig:results_examples1}.

\begin{figure}[htbp]
	\centering
	\includegraphics[width=0.65\textwidth]{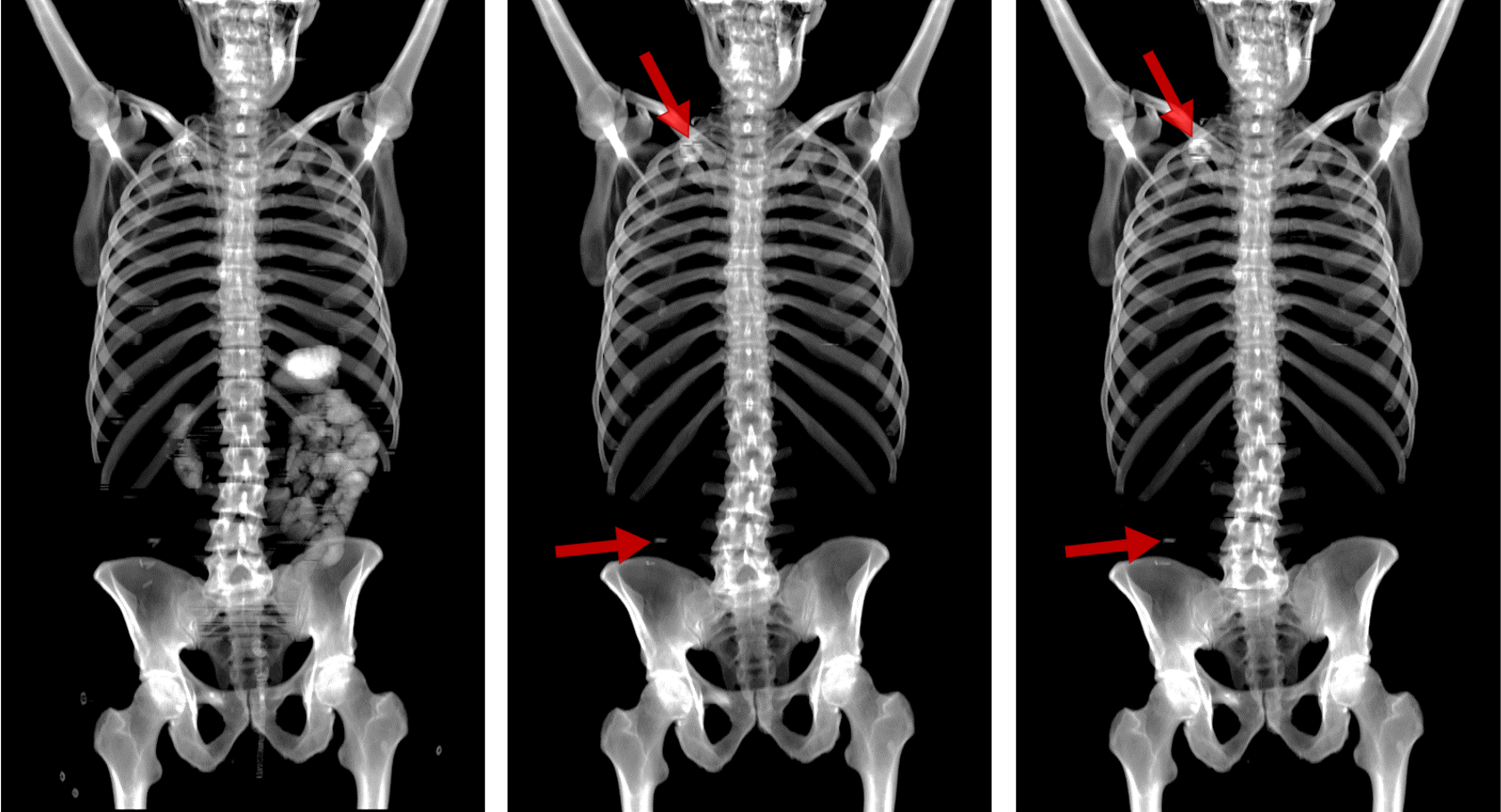}
	\caption{Coronal projections of segmentation outputs of Internal Dataset 1 for hysteresis (\textit{left}), 0 to 255 HU (\textit{centre}), and -100 to 1000 HU (\textit{right}) models. Red arrows indicate segmentation errors.}
	\label{fig:res_exam_int1}
\end{figure}

When compared to hysteresis thresholding shown in Figure \ref{fig:res_exam_int1} both of the U-net based methods produce far superior BBM segmentations. Where hysteresis has several regions of contrast dye throughout the scan, as well as additional metal artifacts, the U-nets achieve close to perfect segmentations across the WB volume.

\subsubsection{Internal Test Dataset 2:}
There were many features present in this dataset that made it more challenging to segment. The main differences between the model performance depended on the degree to which IV and oral contrast were removed in final segmentation. 

\begin{figure}[htbp]
	\centering
	\includegraphics[width=0.85\textwidth]{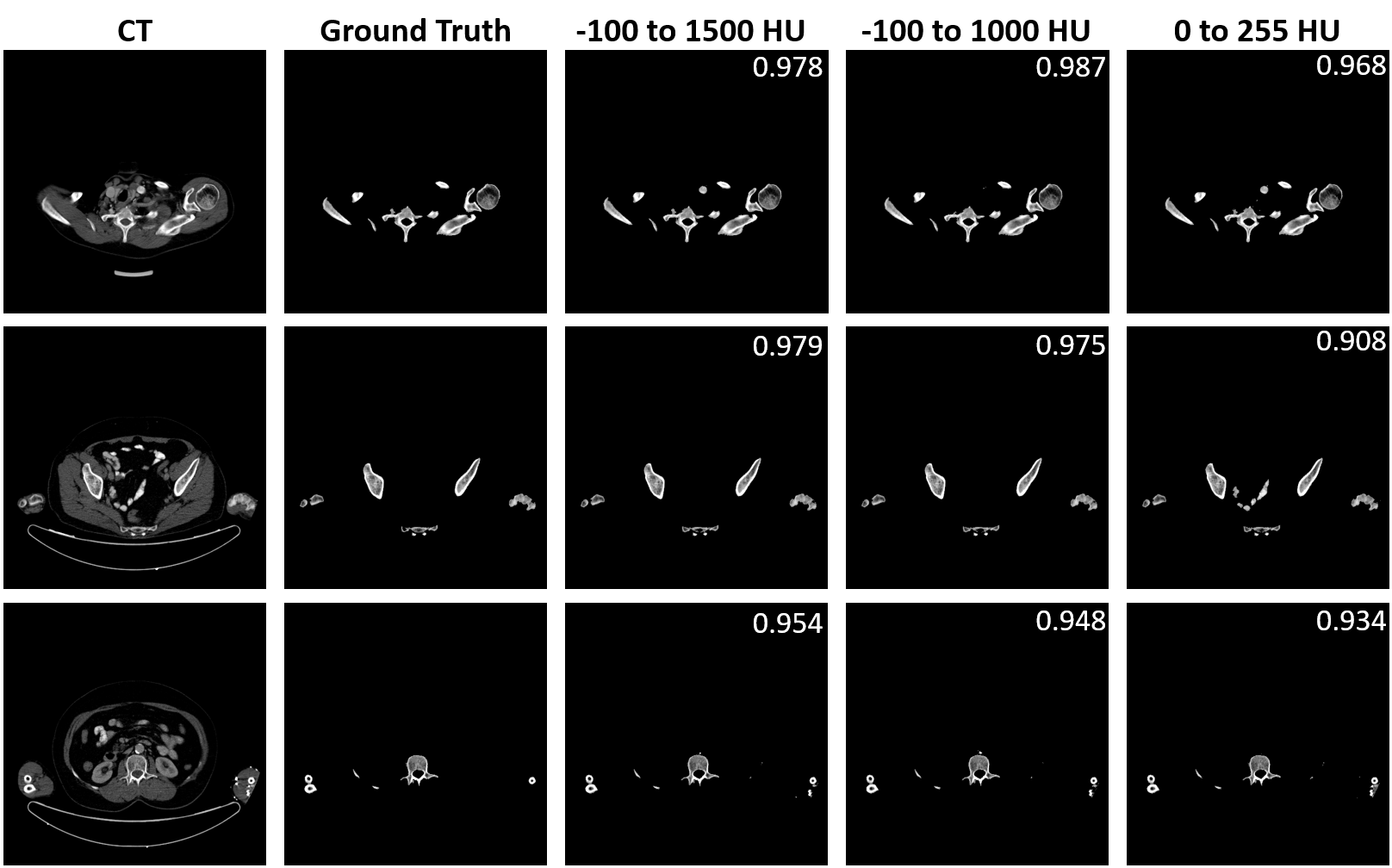}
	\caption{A selection of CT inputs, ground truth, and U-net segmentations demonstrating a selection of DSC results, provided in top right corners, on Internal Test Dataset 2.}
	\label{fig:results_examples2}
\end{figure}

In Figure \ref{fig:results_examples2} (top row) the IV contrast in blood vessels of the neck was only partially removed by U-net models with smaller HU ranges tending to retain more contrast. The majority of models successfully removed contrast in abdominal region with the 0 to 255 performing the least well in this respect with an associated DSC score of 0.908. 

Many of the low DSC scores were associated with the lower spine. It was noted that, within this region, the inclusion of very small areas of false positive or false negative had a greater impact on the DSC due to smaller total overlap available. This is illustrated in Figure \ref{fig:results_examples2} (bottom row), where the difference between the model segmentations and ground truth is minimal and yet the corresponding DSC scores, when considered in isolation, indicate bigger differences between models than observed.

\begin{figure}[htbp]
	\centering
	\includegraphics[width=0.65\textwidth]{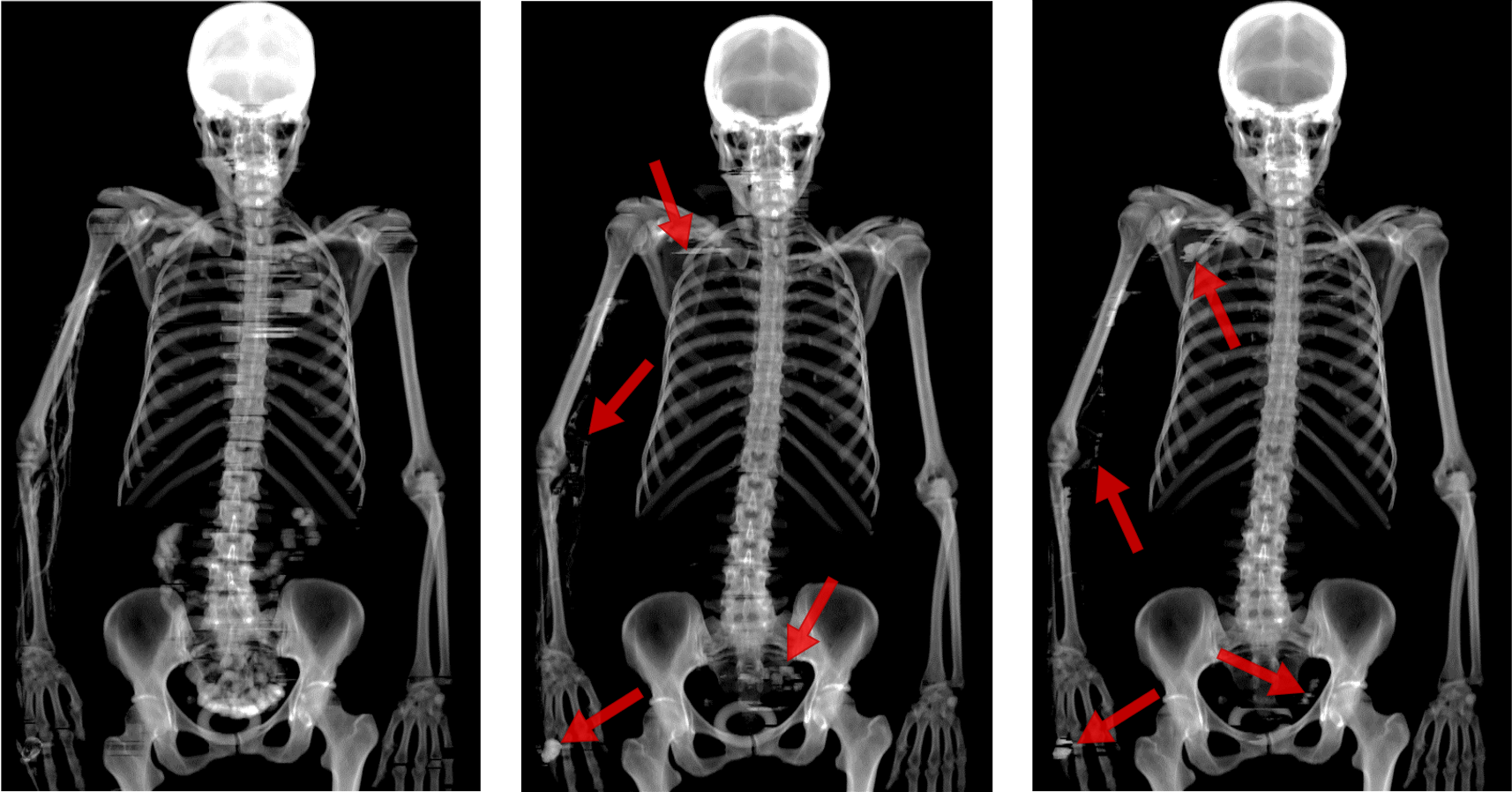}
	\caption{Coronal projections of segmentation outputs of Internal Dataset 2 for hysteresis (\textit{left}), 0 to 255 HU (\textit{centre}), and -100 to 1000 HU (\textit{right}) models. Red arrows indicate segmentation errors.}
	\label{fig:res_exam_int2}
\end{figure}

From the WB projections presented in Figure \ref{fig:res_exam_int2}, it is clear that both of the U-nets are superior to the hysteresis-based approach, most notably with respect to contrast removal. The -100 to 1000 HU ranged model only retained relatively small amounts of IV contrast in arm, and abdo-pelvis region as well as partially removing the pessary device which was almost completely retained by the 0 to 255 HU model.
\subsubsection{External Test Dataset:}
Performance on the external dataset was more varied than on internal data. The best performance was seen when the input CT was close in appearance to images from the training set. 

\begin{figure}[htbp]
	\centering
	\includegraphics[width=0.85\textwidth]{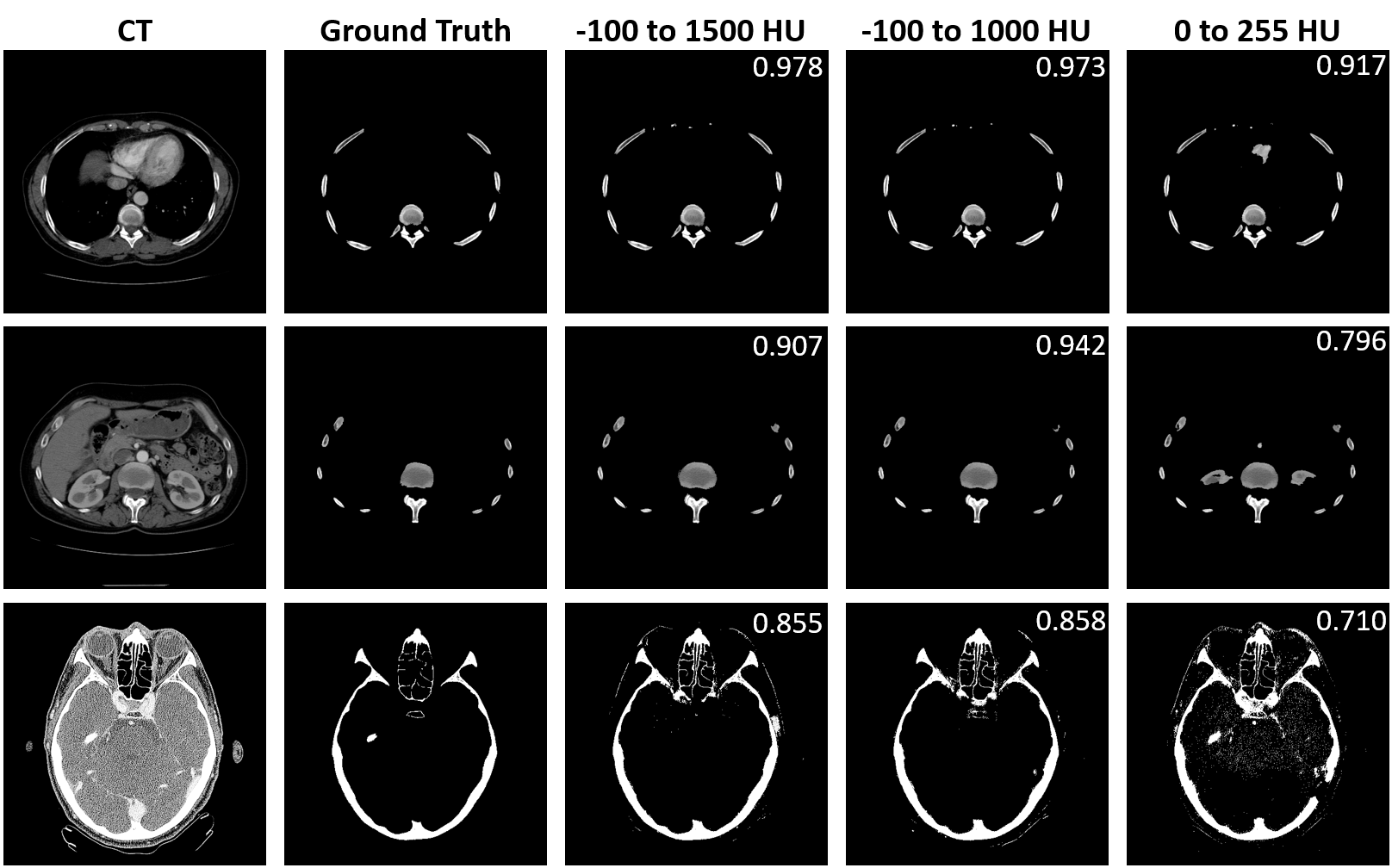}
	\caption{Example inputs, ground truth, and U-net segmentations demonstrating a variety of DSC results, given in top right corners, on the External Dataset.}
	\label{fig:results_examples3}
\end{figure}

Images, such as the small field of view of the head in Figure \ref{fig:results_examples3} (bottom row), were where the lowest DSC scores were observed. The absence of any similar small field images in our training data is a likely factor for this DSC reduction. Images within the dataset were also more varied in appearance in terms of contrast and sharpness than our training/testing data. This appears to have been a confounding factor for the U-nets and introduced a bias in the networks leading to increases in FPs. 

\subsection{Comparison with other studies}
\begin{table}[htbp]
	\caption{Summary of other relevant models from literature.}
	\begin{adjustbox}{max width=\textwidth}
		\begin{tabular}{rcccc}
			\hline
			\multicolumn{1}{l}{}                                   & \textbf{Peréz-Carrasco et al.} \cite{perez2018joint} & \textbf{Klein et al.} \cite{klein2019automatic} & \textbf{Noguchi et al.} \cite{noguchi2020bone} & \textbf{Proposed Model}  \\ \hline
			\textbf{Method}                                & Energy Minimisation                                  & CNN (U-net)                                   & CNN (U-net)                                  & CNN (U-net)           \\
			\textbf{Loss Metric}                                  & \textit{n/a}                                          & Combination of BCE and DSC              & DSC                                    & BCE             \\
			\multicolumn{1}{c}{\textbf{DSC Internal Test Data}} & $0.88 \pm0.14$                              & $0.95\pm0.01$                                  & $0.983 \pm0.005$                        & $0.979\pm0.02$  \\
			&                                     &                                         	&                                        & $0.965\pm0.03$ \\
			\multicolumn{1}{c}{\textbf{Size Internal Dataset}} & $270$                              & $\mathtt{\sim}7200$                                  & $16218$                        &  $1812$ \\
			
			\textbf{DSC External Test Data}                     & \textit{n/a }                                         & $0.92\pm0.05$                                  & $0.961\pm0.007$                         & $0.934\pm0.06$ \\ \hline
		\end{tabular}
	\end{adjustbox}
	\label{table:Comparison}
\end{table}
Table \ref{table:Comparison} presents the DSC ($\pm$ sd) results and additional relevant details from other previous BBM segmentation studies for the purpose of comparison with our presented method. The respective scores relating to internal datasets and the same external dataset are shown separately.

Our -100 to 100 HU ranged model achieved a DSC of 0.979 $\pm$0.02 on  Internal Dataset 1 and 0.965 $\pm$0.03 on Internal Dataset 2 which in combination gives a score of 0.972 $\pm$0.03 across the full internal test data. This is higher than the results of Klein et \textit{et al} \cite{klein2019automatic} whose model was trained/tested on whole body images similar to our own data but with a larger number of images ($n = \mathtt{\sim}7200$). Our results are marginally lower than Noguchi \textit{et al} \cite{noguchi2020bone} whose model trained on various smaller volume scans with a much larger number of images used for training ($n = 16218$). In addition, the dataset of Noguchi \textit{et al} is not reported as a low dose scan and as such this suggests the issues associated with low dose scans discussed in Section \ref{subsec:bckgrnd} may not have been present in this data.

On the external dataset our model performed better than Klein \textit{et al}, but did not achieve the same level of performance reported by Noguchi \textit{et al}. However, both of these studies used the external dataset as part of training or fine-tuning of their models. In contrast we have trained exclusively on the internal data, keeping the external dataset completely separate for the purposes of testing only.
\section{Discussion}
In this paper, we have proposed a novel approach to BBM segmentation in CE low dose WB-CT through the application of additional preprocessing to the training data designed to enhance a models ability to successfully differentiate between high density regions due to bone and contrast dye. In addition we have introduced an analytical means of determining the threshold value of the sigmoid activation output of the U-net.

Analysis of test datasets with moderate to high levels of contrast has demonstrated that our method is effective in producing accurate WB BBM segmentations in CE CT.

To the best of our knowledge, our research is the first in the literature where all of our data are CE low dose WB-CT scans. This has allowed for an in-depth characterisation of the complexity that contrast dye poses for BBM segmentation in addition to a novel solution to be presented using a much smaller dataset than other similar studies (see Table \ref{table:Comparison}).

We have assessed five training/testing datasets each of which captured different amounts of BBM contrast detail. This was implemented as an additional preprocessing step through adjustment of the HU range prior to image conversion from DICOM to PNG format required for CNN development in Keras Tensorflow. We have found that for CE scans when larger ranges, such as -100 to 1500 HU and -100 to 1000 HU, were used the U-net was more successful in identifying and excluding contrast dye in BBM segmentations on a challenging dataset.

Our results, see Figure \ref{fig:dsc_images}, agree with similar WB data of Klein \textit{et al} \cite{klein2019automatic} in demonstrating the tendency of DSC scores to fluctuate across patient volumes. As such, the use of ANOVA on the logit (DSC) is a useful tool for assessing differences between models in comparison to sole reliance on average DSC scores.

We have demonstrated the ability of our models to generalize to a new data despite significant differences between training and external datasets previously discussed. However, it is not clear whether or not the subjects in the dataset of Peréz-Carrasco \textit{et al} had underlying conditions that would impact the appearance of BBM in the images. The scans used in our study were restricted to healthy BBM patients. It is therefore not possible to know how well the system will generalize to patients suffering from the conditions discussed in Section \ref{subsec:bckgrnd} and in other similar studies \cite{noguchi2020bone,klein2019automatic} where bone appearance may be altered.

Although our original datasets were PET-CT, we utilised only the CT component in the research presented. The methods we have outlined can be applied with little modification to the PET component of the scan, and improve accuracy of previous bone metabolism assessment tools which are reliant on artifact prone thresholding approaches \cite{leydon2017semi,takahashi2019proposal}.

\section{Conclusions}

We have outlined a U-net deep learning architecture with novel preprocessing techniques to successfully segment BBM regions from low dose CE WB-CT scans. We have demonstrated that, when wider ranges of HU were used for the training and testing data, the performance of CNNs improved in terms of differentiating between bone and contrast dye. We have also shown that excellent results can be achieved using comparatively small datasets ($n=1812$) comprised of low dose CT scans.

%There were only marginal differences observed between models trained using variations of the BCE loss function. Significant differences were observed when testing on an External dataset where higher values of the $\beta$ weighting performed better than lower values, and the standard BCE loss suggesting an improved generalization ability.

\section*{References}
%\begin{thebibliography}{10}
\bibliographystyle{vancouver}
\bibliography{DRAFT_Segmentation_Paper}

\begin{thebibliography}{10}

\bibitem{hillengass2019international}
Hillengass J, Usmani S, Rajkumar SV, Durie BG, Mateos MV, Lonial S, et~al.
\newblock International myeloma working group consensus recommendations on
  imaging in monoclonal plasma cell disorders.
\newblock The Lancet Oncology. 2019;20(6):e302--e312.

\bibitem{hoilund2018cancer}
H{\o}ilund-Carlsen PF, Hess S, Werner TJ, Alavi A. Cancer metastasizes to the
  bone marrow and not to the bone: time for a paradigm shift!
\newblock Springer; 2018.

\bibitem{macedo2017bone}
Macedo F, Ladeira K, Pinho F, Saraiva N, Bonito N, Pinto L, et~al.
\newblock Bone metastases: an overview.
\newblock Oncology reviews. 2017;11(1).

\bibitem{kim2019visual}
Kim I, Rajaraman S, Antani S.
\newblock Visual interpretation of convolutional neural network predictions in
  classifying medical image modalities.
\newblock Diagnostics. 2019;9(2):38.

\bibitem{gordon2008automated}
Gordon L, Hardisty M, Skrinskas T, Wu F, Whyne C.
\newblock Automated atlas-based 3D segmentation of the metastatic spine.
\newblock In: Orthopaedic Proceedings. vol.~90. The British Editorial Society
  of Bone \& Joint Surgery; 2008. p. 128--128.

\bibitem{boehm1999three}
Boehm G, Knoll CJ, Colomer VG, Alcaniz-Raya ML, Albalat SE.
\newblock Three-dimensional segmentation of bone structures in CT images.
\newblock In: Medical Imaging 1999: Image Processing. vol. 3661. International
  Society for Optics and Photonics; 1999. p. 277--286.

\bibitem{pinheiro2015new}
Pinheiro M, Alves J.
\newblock A new level-set-based protocol for accurate bone segmentation from CT
  imaging.
\newblock IEEE Access. 2015;3:1894--1906.

\bibitem{krvcah2011fully}
Kr{\v{c}}ah M, Sz{\'e}kely G, Blanc R.
\newblock Fully automatic and fast segmentation of the femur bone from 3D-CT
  images with no shape prior.
\newblock In: 2011 IEEE international symposium on biomedical imaging: from
  nano to macro. IEEE; 2011. p. 2087--2090.

\bibitem{burdin1994surface}
Burdin V, Roux C.
\newblock Surface segmentation of long bone structures from 3D CT images using
  a deformable contour model.
\newblock In: Proceedings of 16th Annual International Conference of the IEEE
  Engineering in Medicine and Biology Society. vol.~1. IEEE; 1994. p. 512--513.

\bibitem{burnett2004deformable}
Burnett SS, Starkschall G, Stevens CW, Liao Z.
\newblock A deformable-model approach to semi-automatic segmentation of CT
  images demonstrated by application to the spinal canal.
\newblock Medical physics. 2004;31(2):251--263.

\bibitem{franzle2014fully}
Fr{\"a}nzle A, Sumkauskaite M, Hillengass J, B{\"a}uerle T, Bendl R.
\newblock Fully automated shape model positioning for bone segmentation in
  whole-body CT scans.
\newblock In: Journal of Physics: Conference Series. vol. 489. IOP Publishing;
  2014. p. 012029.

\bibitem{natsheh2010segmentation}
Natsheh AR, Ponnapalli PV, Anani N, Benchebra D, El-kholy A, Norburn P.
\newblock Segmentation of bone structure in sinus CT images using
  self-organizing maps.
\newblock In: 2010 IEEE International Conference on Imaging Systems and
  Techniques. IEEE; 2010. p. 294--299.

\bibitem{guo20183d}
Guo H, Song S, Wang J, Guo M, Cheng Y, Wang Y, et~al.
\newblock 3D surface voxel tracing corrector for accurate bone segmentation.
\newblock International journal of computer assisted radiology and surgery.
  2018;13(10):1549--1563.

\bibitem{sharma2010automated}
Sharma N, Aggarwal LM.
\newblock Automated medical image segmentation techniques.
\newblock Journal of medical physics/Association of Medical Physicists of
  India. 2010;35(1):3.

\bibitem{puri2012semiautomatic}
Puri T, Blake GM, Curran KM, Carr H, Moore AE, Colgan N, et~al.
\newblock Semiautomatic region-of-interest validation at the femur in
  18f-fluoride pet/ct.
\newblock Journal of nuclear medicine technology. 2012;40(3):168--174.

\bibitem{belal2019deep}
Belal SL, Sadik M, Kaboteh R, Enqvist O, Ul{\'e}n J, Poulsen MH, et~al.
\newblock Deep learning for segmentation of 49 selected bones in CT scans:
  First step in automated PET/CT-based 3D quantification of skeletal
  metastases.
\newblock European journal of radiology. 2019;113:89--95.

\bibitem{klein2019automatic}
Klein A, Warszawski J, Hillenga{\ss} J, Maier-Hein KH.
\newblock Automatic bone segmentation in whole-body CT images.
\newblock International journal of computer assisted radiology and surgery.
  2019;14(1):21--29.

\bibitem{sanchez2020segmentation}
S{\'a}nchez JCG, Magnusson M, Sandborg M, Tedgren {\AA}C, Malusek A.
\newblock Segmentation of bones in medical dual-energy computed tomography
  volumes using the 3D U-Net.
\newblock Physica Medica. 2020;69:241--247.

\bibitem{noguchi2020bone}
Noguchi S, Nishio M, Yakami M, Nakagomi K, Togashi K.
\newblock Bone segmentation on whole-body CT using convolutional neural network
  with novel data augmentation techniques.
\newblock Computers in Biology and Medicine. 2020;p. 103767.

\bibitem{boehm2016physics}
Boehm IB, Heverhagen JT.
\newblock Physics of computed tomography: contrast agents.
\newblock In: Handbook of neuro-oncology neuroimaging. Elsevier; 2016. p.
  151--155.

\bibitem{lusic2013x}
Lusic H, Grinstaff MW.
\newblock X-ray-computed tomography contrast agents.
\newblock Chemical reviews. 2013;113(3):1641--1666.

\bibitem{fiebich1999automatic}
Fiebich M, Straus CM, Sehgal V, Renger BC, Doi K, Hoffmann KR.
\newblock Automatic bone segmentation technique for CT angiographic studies.
\newblock Journal of computer assisted tomography. 1999;23(1):155--161.

\bibitem{kalra2019contrast}
Kalra MK, Becker HC, Enterline DS, Lowry CR, Molvin LZ, Singh R, et~al.
\newblock Contrast administration in CT: a patient-centric approach.
\newblock Journal of the American College of Radiology. 2019;16(3):295--301.

\bibitem{goodfellow2016deep2}
Goodfellow I, Bengio Y, Courville A.
\newblock Deep learning.
\newblock MIT press; 2016.

\bibitem{litjens2017survey}
Litjens G, Kooi T, Bejnordi BE, Setio AAA, Ciompi F, Ghafoorian M, et~al.
\newblock A survey on deep learning in medical image analysis.
\newblock Medical image analysis. 2017;42:60--88.

\bibitem{ronneberger2015u}
Ronneberger O, Fischer P, Brox T.
\newblock U-net: Convolutional networks for biomedical image segmentation.
\newblock In: International Conference on Medical image computing and
  computer-assisted intervention. Springer; 2015. p. 234--241.

\bibitem{ciresan2012deep}
Ciresan D, Giusti A, Gambardella LM, Schmidhuber J.
\newblock Deep neural networks segment neuronal membranes in electron
  microscopy images.
\newblock In: Advances in neural information processing systems; 2012. p.
  2843--2851.

\bibitem{drozdzal2016importance}
Drozdzal M, Vorontsov E, Chartrand G, Kadoury S, Pal C.
\newblock The importance of skip connections in biomedical image segmentation.
\newblock In: Deep Learning and Data Labeling for Medical Applications.
  Springer; 2016. p. 179--187.

\bibitem{iglovikov2017satellite}
Iglovikov V, Mushinskiy S, Osin V.
\newblock Satellite imagery feature detection using deep convolutional neural
  network: A kaggle competition.
\newblock arXiv preprint arXiv:170606169. 2017;.

\bibitem{kazemifar2018segmentation}
Kazemifar S, Balagopal A, Nguyen D, McGuire S, Hannan R, Jiang S, et~al.
\newblock Segmentation of the prostate and organs at risk in male pelvic CT
  images using deep learning.
\newblock Biomedical Physics \& Engineering Express. 2018;4(5):055003.

\bibitem{iglovikov2018paediatric}
Iglovikov VI, Rakhlin A, Kalinin AA, Shvets AA.
\newblock Paediatric bone age assessment using deep convolutional neural
  networks.
\newblock In: Deep Learning in Medical Image Analysis and Multimodal Learning
  for Clinical Decision Support. Springer; 2018. p. 300--308.

\bibitem{zeng20173d}
Zeng G, Yang X, Li J, Yu L, Heng PA, Zheng G.
\newblock 3D U-net with multi-level deep supervision: fully automatic
  segmentation of proximal femur in 3D MR images.
\newblock In: International workshop on machine learning in medical imaging.
  Springer; 2017. p. 274--282.

\bibitem{christ2017automatic}
Christ PF, Ettlinger F, Gr{\"u}n F, Elshaera MEA, Lipkova J, Schlecht S, et~al.
\newblock Automatic liver and tumor segmentation of CT and MRI volumes using
  cascaded fully convolutional neural networks.
\newblock arXiv preprint arXiv:170205970. 2017;.

\bibitem{milletari2016v}
Milletari F, Navab N, Ahmadi SA.
\newblock V-net: Fully convolutional neural networks for volumetric medical
  image segmentation.
\newblock In: 2016 Fourth International Conference on 3D Vision (3DV). IEEE;
  2016. p. 565--571.

\bibitem{cciccek20163d}
{\c{C}}i{\c{c}}ek {\"O}, Abdulkadir A, Lienkamp SS, Brox T, Ronneberger O.
\newblock 3D U-Net: learning dense volumetric segmentation from sparse
  annotation.
\newblock In: International conference on medical image computing and
  computer-assisted intervention. Springer; 2016. p. 424--432.

\bibitem{suzuki2010computer}
Suzuki K, Kohlbrenner R, Epstein ML, Obajuluwa AM, Xu J, Hori M.
\newblock Computer-aided measurement of liver volumes in CT by means of
  geodesic active contour segmentation coupled with level-set algorithms.
\newblock Medical physics. 2010;37(5):2159--2166.

\bibitem{tappeiner2019multi}
Tappeiner E, Pr{\"o}ll S, H{\"o}nig M, Raudaschl PF, Zaffino P, Spadea MF,
  et~al.
\newblock Multi-organ segmentation of the head and neck area: an efficient
  hierarchical neural networks approach.
\newblock International journal of computer assisted radiology and surgery.
  2019;14(5):745--754.

\bibitem{alirr2018automated}
Alirr OI, Rahni AAA, Golkar E.
\newblock An automated liver tumour segmentation from abdominal CT scans for
  hepatic surgical planning.
\newblock International journal of computer assisted radiology and surgery.
  2018;13(8):1169--1176.

\bibitem{canny1983finding}
Canny JF.
\newblock Finding edges and lines in images.
\newblock MASSACHUSETTS INST OF TECH CAMBRIDGE ARTIFICIAL INTELLIGENCE LAB;
  1983.

\bibitem{sogo2012assessment}
Sogo M, Ikebe K, Yang TC, Wada M, Maeda Y.
\newblock Assessment of bone density in the posterior maxilla based on
  Hounsfield units to enhance the initial stability of implants.
\newblock Clinical implant dentistry and related research. 2012;14:e183--e187.

\bibitem{eddins2004digital}
Eddins SL, Gonzalez R, Woods R.
\newblock Digital image processing using Matlab.
\newblock Princeton Hall Pearson Education Inc, New Jersey. 2004;.

\bibitem{MATLAB:2018b}
{MATLAB version 9.5 (R2018b)}.
\newblock Natick, Massachusetts; 2018.

\bibitem{zhu2012automatic}
Zhu YM, Cochoff SM, Sukalac R.
\newblock Automatic patient table removal in CT images.
\newblock Journal of digital imaging. 2012;25(4):480--485.

\bibitem{bandi2011automated}
Bandi P, Zsoter N, Seres L, Toth Z, Papp L.
\newblock Automated patient couch removal algorithm on CT images.
\newblock In: 2011 Annual International Conference of the IEEE Engineering in
  Medicine and Biology Society. IEEE; 2011. p. 7783--7786.

\bibitem{baron1994understanding}
Baron R.
\newblock Understanding and optimizing use of contrast material for CT of the
  liver.
\newblock AJR American journal of roentgenology. 1994;163(2):323--331.

\bibitem{barrett2004artifacts}
Barrett JF, Keat N.
\newblock Artifacts in CT: recognition and avoidance.
\newblock Radiographics. 2004;24(6):1679--1691.

\bibitem{willemink2020preparing}
Willemink MJ, Koszek WA, Hardell C, Wu J, Fleischmann D, Harvey H, et~al.
\newblock Preparing medical imaging data for machine learning.
\newblock Radiology. 2020;295(1):4--15.

\bibitem{perez2018joint}
P{\'e}rez-Carrasco JA, Acha B, Su{\'a}rez-Mej{\'\i}as C, L{\'o}pez-Guerra JL,
  Serrano C.
\newblock Joint segmentation of bones and muscles using an intensity and
  histogram-based energy minimization approach.
\newblock Computer methods and programs in biomedicine. 2018;156:85--95.

\bibitem{nair2010rectified}
Nair V, Hinton GE.
\newblock Rectified linear units improve restricted boltzmann machines.
\newblock In: Proceedings of the 27th international conference on machine
  learning (ICML-10); 2010. p. 807--814.

\bibitem{kingma2014adam}
Kingma DP, Ba J.
\newblock Adam: A method for stochastic optimization.
\newblock arXiv preprint arXiv:14126980. 2014;.

\bibitem{reddi2019convergence}
Reddi SJ, Kale S, Kumar S.
\newblock On the convergence of adam and beyond.
\newblock arXiv preprint arXiv:190409237. 2019;.

\bibitem{feng2017discriminative}
Feng X, Yang J, Laine AF, Angelini ED.
\newblock Discriminative localization in CNNs for weakly-supervised
  segmentation of pulmonary nodules.
\newblock In: International Conference on Medical Image Computing and
  Computer-Assisted Intervention. Springer; 2017. p. 568--576.

\bibitem{ibtehaz2020multiresunet}
Ibtehaz N, Rahman MS.
\newblock MultiResUNet: Rethinking the U-Net architecture for multimodal
  biomedical image segmentation.
\newblock Neural Networks. 2020;121:74--87.

\bibitem{leger2019deep}
L{\'e}ger J, Leyssens L, De~Vleeschouwer C, Kerckhofs G.
\newblock Deep learning-based segmentation of mineralized cartilage and bone in
  high-resolution micro-CT images.
\newblock In: International Symposium on Computer Methods in Biomechanics and
  Biomedical Engineering. Springer; 2019. p. 158--170.

\bibitem{ding2019votenet}
Ding Z, Han X, Niethammer M.
\newblock VoteNet: a deep learning label fusion method for multi-atlas
  segmentation.
\newblock In: International Conference on Medical Image Computing and
  Computer-Assisted Intervention. Springer; 2019. p. 202--210.

\bibitem{dice1945measures}
Dice LR.
\newblock Measures of the amount of ecologic association between species.
\newblock Ecology. 1945;26(3):297--302.

\bibitem{taha2015metrics}
Taha AA, Hanbury A.
\newblock Metrics for evaluating 3D medical image segmentation: analysis,
  selection, and tool.
\newblock BMC medical imaging. 2015;15(1):29.

\bibitem{mathworks2019statistics}
Mathworks.
\newblock Statistics and Machine Learning Toolbox™ User’s Guide (R 2019b).
  2019;p. 409.

\bibitem{zou2004statistical}
Zou KH, Warfield SK, Bharatha A, Tempany CM, Kaus MR, Haker SJ, et~al.
\newblock Statistical validation of image segmentation quality based on a
  spatial overlap index1: scientific reports.
\newblock Academic radiology. 2004;11(2):178--189.

\bibitem{leydon2017semi}
Leydon P, O’Connell M, Greene D, Curran K.
\newblock Semi-automatic Bone Marrow Evaluation in PETCT for Multiple Myeloma.
\newblock In: Annual Conference on Medical Image Understanding and Analysis.
  Springer; 2017. p. 342--351.

\bibitem{takahashi2019proposal}
Takahashi ME, Mosci C, Souza EM, Brunetto SQ, Etchebehere E, Santos AO, et~al.
\newblock Proposal for a Quantitative 18 F-FDG PET/CT Metabolic Parameter to
  Assess the Intensity of Bone Involvement in Multiple Myeloma.
\newblock Scientific reports. 2019;9(1):1--8.

\end{thebibliography}
%\end{thebibliography}.

\end{document}